\documentclass[twocolumn,aps,superscriptaddress,nofootinbib]{revtex4-1}
\usepackage[dvipsnames]{xcolor}
\usepackage{amsmath}
\usepackage{amsfonts}
\usepackage{amssymb}
\usepackage{hyperref}
\usepackage{graphicx}
\usepackage{bm}

\begin{document}
\title{Scheme for Deterministic Loading of Laser-Cooled Molecules into Optical Tweezers}

\author{Etienne F. Walraven}
\affiliation{Institute for Molecules and Materials, Radboud University, Nijmegen, The Netherlands}
\author{Michael R. Tarbutt}
\affiliation{Centre for Cold Matter, Blackett Laboratory, Imperial College London, Prince Consort Road, London SW7 2AZ, United Kingdom}
\author{Tijs Karman}
\email{tkarman@science.ru.nl}
\affiliation{Institute for Molecules and Materials, Radboud University, Nijmegen, The Netherlands}

\begin{abstract}
We propose to repeatedly load laser-cooled molecules into optical tweezers,
and transfer them to storage states that are rotationally excited by two additional quanta.
Collisional loss of molecules in these storage states is suppressed,
and a dipolar blockade prevents the accumulation of more than one molecule.
Applying three cycles loads tweezers with single molecules at an 80~\% success rate, limited by residual collisional loss.
This improved loading efficiency reduces the time needed for rearrangement of tweezer arrays, which would otherwise limit the scalability of neutral molecule quantum computers.
\end{abstract}

\maketitle

Optical tweezer arrays of atoms and molecules have emerged as powerful platforms for quantum information processing \cite{kaufman:21}.
Interactions between atoms in different tweezers can be induced by excitation to Rydberg states which can be used to implement fast two-qubit gates.
Tweezer arrays of ultracold molecules have recently also been realized \cite{anderegg:19}.
Polar molecules give access to tunable dipole-dipole interactions \cite{carr:09} in the electronic ground state that can be used to implement quantum gates \cite{yelin:06,ni:18,holland:23,bao:23} without losses by spontaneous emission or anti-trapping in the interacting state.
Tweezers can be loaded from dilute gases, foregoing the need to reach high phase-space density, which remains challenging for ultracold molecules \cite{li:21,anderegg:21,schindewolf:22,lin:23,bigagli:23}.

Tweezers are typically loaded from a gas of laser-cooled atoms or molecules. 
Laser cooling provides the dissipation needed to load the conservative potential of the tweezer.
When a second particle is loaded in the $\mu$m-sized tweezer, the pair is quickly lost by light-assisted collisions. Thus, a tweezer contains zero or one particle, each with approximately 50~\% probability, depending on whether an even or odd number was loaded in total. This is referred to as collisional blockade \cite{schlosser:02}.
Current tweezer arrays for molecules are linear arrays of typically 20-40 sites with waist size of 730 nm and a loading probability of 30-40~\% \cite{anderegg:19,holland:23,bao:23}. In these experiments, the temperature was about 100~$\mu$K, but methods to cool to a few~$\mu$K are well established \cite{cheuk:18,caldwell:19} and molecules have recently been cooled to the ground state of motion of the tweezer \cite{lu:24,bao:24}.

A powerful feature of tweezer arrays is the ability to rearrange the stochastically-loaded half-filled array into smaller defect-free arrays \cite{kaufman:21}.
However, the trapped particles have a limited lifetime and the time taken for rearrangement increases with array size, ultimately limiting the scalability of this approach.
To circumvent this, \emph{deterministic} loading of tweezers with single particles is desirable.
For ultracold atoms, near deterministic loading has been demonstrated using two different methods. 
In the first method, discussed in more detail below, the light-assisted collision is controlled \cite{grunzweig:10,brown:19} so that only one of the two atoms is ejected, yielding a success probability of about 90~\%. 
In the second method, Sr atoms were loaded with 84~\% probability by imaging loaded sites,
then shelving those atoms into dark states and lowering their trap depths to prevent new atoms loading into the same sites on the next iteration \cite{shaw:23}.
This scheme requires active feedback and can in principle be extended to molecules~\cite{shaw:23}.

\begin{figure}
\centering
\includegraphics[width=0.45\textwidth]{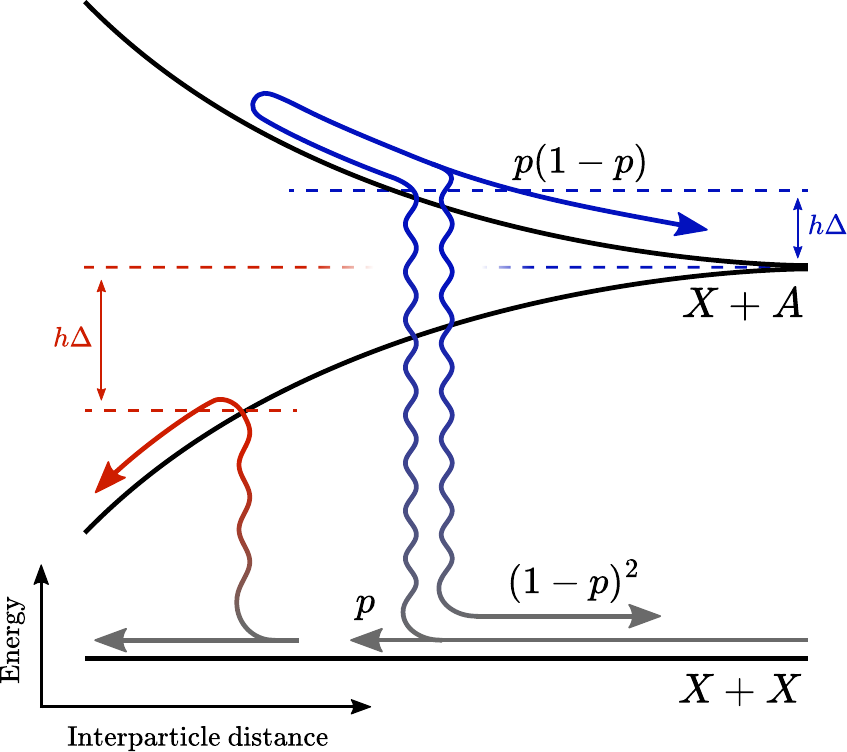}
\caption{
{\bf Light-assisted collisions}.
Shown schematically are interaction potentials in the electronic ground state $X$ and excited state $A$.
Wavy vertical lines indicate the Condon point where the attractive or repulsive dipole-dipole interaction in the excited state compensates the red or blue detuning, respectively.
Non-adiabatic transitions occur at the Condon point with probability $p$, leading to a short range encounter. For blue detuning, there can be a light-assisted collision on the repulsive branch, releasing kinetic energy $h\Delta$ which can be controlled to eject only one of the two particles. This requires a process that is adiabatic on the way in, and non-adiabatic on the way out, so has probability $p(1-p)$ whose maximum value is 0.25. For atoms, short-range encounters are usually elastic, so the atoms have many opportunities for such controlled light-assisted collisions. For molecules, short-range encounters usually lead to loss, and the loss dominates since $p$ is always larger than $p(1-p)$.
\label{fig:zero}
}
\end{figure}

\begin{figure*}
\centering
\includegraphics[width=0.95\textwidth]{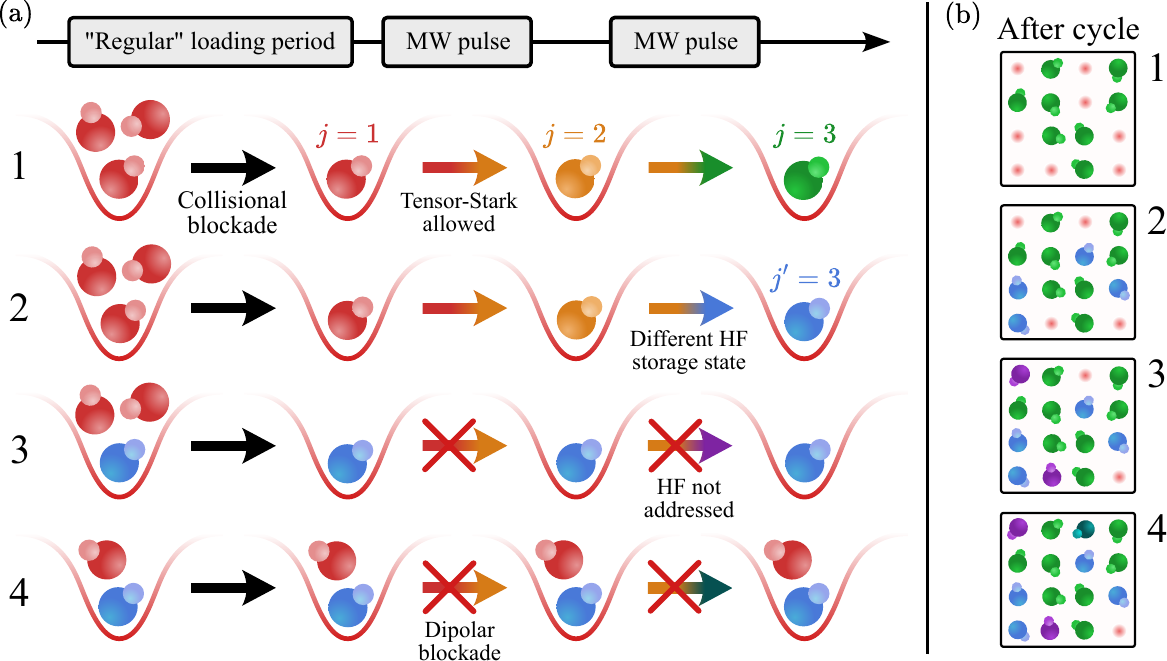}
\caption{
{\bf Proposed scheme for deterministic loading of laser-cooled molecules into optical tweezers.}
({\bf a}) Illustrates the different processes that can occur in each tweezer in each cycle.
(Row~{\bf 1}) Single $j=1$ molecules are loaded with 50~\% probability due to collisional blockade. They are transferred to $j=3$ using two microwave $\pi$ pulses.
(Row~{\bf 2}) Subsequent cycles transfer molecules to different hyperfine states of $j=3$, indicated by different colors.
(Row~{\bf 3}) Molecules loaded in previous cycles are not addressed by subsequent microwave pulses and remain stored.
(Row~{\bf 4}) When a tweezer already contains a $j=3$ molecule, the dipolar blockade prevents accumulating additional $j=3$ molecules.
({\bf b}) Illustrates loading of the array after multiple cycles of the scheme.
\label{fig:scheme}
}
\end{figure*}

The established picture \cite{brown:19,grunzweig:10} of deterministic loading by light-assisted collisions is illustrated in Fig.~\ref{fig:zero}.
Non-adiabatic transitions can occur at the Condon point where the dipole-dipole interaction in the excited state compensates the detuning.
In the excited state, loss can occur either by spontaneous emission at a different interatomic distance,
or by dissociation in the excited state for blue detuning.
By choosing the detuning appropriately, one can ensure collisions can remove one, but never two atoms from the tweezer.
Removal of a single atom requires an uneven distribution of the kinetic energy release, which does not occur for two atoms initially at rest.
Thus for atoms cooled to the bottom of the tweezer,
single-atom removal may require multiple light-assisted collisions,
or elastic collisions, to redistribute the energy release unevenly before the atoms are laser-cooled again.

It does not seem possible to extend this scheme to molecules.
Molecule pairs suffer from fast collisional loss \cite{ospelkaus:10,takekoshi:14, molony:14,guo:16,park:15,gregory:19,cheuk:20} that occurs in short-range encounters \cite{idziaszek:10}.
As illustrated in Fig.~\ref{fig:zero}, the ratio of light-assisted collisions to short-range encounters depends on the probability of non-adiabatic transitions at the Condon point, but the ratio can never exceed 1. 
Collisional shielding \cite{gonzalez:17,karman:18d,lassabliere:18,karam:23} can suppress loss and enable elastic scattering of molecules, required for deterministic loading, but these shielding schemes seem incompatible with laser cooling that is needed for tweezer loading. 
The shielding schemes set up long-range repulsive interactions between molecules prepared in a specific internal state, typically a state of mixed parity, but laser cooling usually relies on parity selection rules and cycling between various substates, some of which have no collisional shielding or even an enhancement of collisional loss \cite{anderegg:21}.

Figure~\ref{fig:scheme} illustrates our scheme for near-deterministic loading of ultracold molecules into tweezers. The scheme combines several concepts: (i) collisions prevent the loading of more than one molecule; (ii) loaded molecules can be transferred to other rotational states where collisions are strongly suppressed; (iii) due to the tensor part of the AC Stark shift, trapped molecules can be distinguished from those in the bulk gas; (iv) stored molecules induce a dipolar blockade that distinguishes loaded tweezers from empty ones. Starting from a laser-cooled sample, we load $j=1$ molecules in the collisionally blockaded regime \cite{schlosser:02}, resulting in a tweezer occupancy of zero or one, each with 50~\% probability. This loading step should leave molecules in a single hyperfine and Zeeman sub-level. Deep laser cooling schemes ~\cite{cheuk:18, caldwell:19} rely on coherent population trapping in a zero-velocity dark state, so automatically pump molecules into a single state. Alternatively, a short pulse of optical pumping can be used to transfer all molecules to a desired component of $j=1$. The loaded molecules are then transferred to a $j=3$ ``storage state'' using two microwave $\pi$-pulses. The transition frequency is shifted by the tensor AC Stark shift of the tweezer, so molecules in the tweezer can be transferred without affecting molecules in the bulk gas. These two steps are applied repeatedly, using a different hyperfine component of $j=3$ for storage in each cycle. 

At the end of each loading period there are four possible outcomes for each tweezer. (1) The tweezer contains only a $j=1$ molecule. It will be transferred to $j=3$. (2) The tweezer contains only a $j=3$ molecule. The molecule will remain in $j=3$ because it is in a different hyperfine component to the one addressed by the microwave field. (3) The tweezer contains both a $j=1$ molecule and a $j=3$ molecule. In this case, the $j=1 \leftrightarrow j=2$ transition is shifted out of resonance with the first $\pi$-pulse due to the dipole-dipole interaction with the stored $j=3$ molecule, and neither molecule is affected by the microwave fields. This is the dipolar blockade. Furthermore, the $j=3$ molecules are protected from collisions with $j=1$ molecules by repulsive van der Waals interactions \cite{walraven:24a}. (4) The tweezer is empty. There is a $50~\%$ chance it will be loaded in the next cycle. After several cycles, the occupancy will be close to 100~\%. Then, any molecules in $j=1$ are removed by resonant light and $j=3$ molecules are transferred back to $j=1$ with two $\pi$ pulses. They will be in different hyperfine states but can easily be transferred to a single state by a final step of optical pumping.

Our scheme is applicable to all the molecular species laser-cooled so far. To evaluate its potential, we focus on the specific example of CaF molecules cooled on the $A ^{2}\Pi_{1/2} - X ^{2}\Sigma$ transition. Throughout, we assume a temperature $T=5$~$\mu$K \cite{cheuk:18, caldwell:19}, and tweezers with wavelength 1064~nm, a depth of 250~$\mu$K and axial and
radial trapping frequencies of $\omega_z/(2\pi) = 7$~kHz and $\omega_r/(2\pi) = 50$~kHz similar to recent experiments \cite{holland:23,bao:23}. 
For these parameters, the peak density is $1.5\times10^{13}$~cm$^{-3}$ and the distribution is
\begin{align}
n(\bm{r}_1) &= \frac{1}{(2\pi)^{3/2}\sigma_x\sigma_y\sigma_z} \exp\left(-\frac{x_1^2}{2\sigma_x^2}\right)\nonumber \\
\times& \exp\left(-\frac{y_1^2}{2\sigma_y^2}\right)\exp\left(-\frac{z_1^2}{2\sigma_z^2}\right)
\end{align}
where $\sigma_i = (k_BT / M \omega_i^2)^{1/2}$, $M$ is the molecule mass, and $\omega_i$ the harmonic trapping frequency of the tweezer in direction $i$.
For a pair of molecules, we use $n(\bm{r}_1)n(\bm{r}_2) = n(\bm{R}\sqrt{2}) n(\bm{r}/\sqrt{2})$ where $\bm{R} = (\bm{r}_1+\bm{r}_2)/2$ is the center of mass coordinate and $\bm{r} = \bm{r}_1-\bm{r}_2$ is the relative coordinate. 
We assume a tweezer loading rate of 100 molecules per second \cite{schlosser:02}, so choose loading periods of 10~ms. This is much longer than the timescale for deep laser cooling~\cite{caldwell:19}. The rotational transition dipole moments are assumed to be $d=\mu_{e}/\sqrt{3}$ where $\mu_{e} = 3.1$~Debye is the molecule-frame dipole moment. 

To evaluate the strength of the dipolar blockade, we calculate the dipole-dipole interaction energy
\begin{align}
U_{dd} &= \iiint \iiint n(\bm{r}_1) V_{dd}(\bm{r}_1-\bm{r}_2) n(\bm{r}_2)~\mathrm{d}\bm{r}_1~\mathrm{d}\bm{r}_2 \nonumber \\
 &= \frac{1}{2^{3/2}} \iiint n(\bm{r}) V_{dd}(\bm{r})~\mathrm{d}\bm{r}.
\end{align}
Here, $V_{dd}$ is the interaction potential which we take as
\begin{align}
V_{dd}(\bm{r}) = \frac{d^2}{4\pi\epsilon_0 r^3} 2P_2(\hat{\bm{r}} \cdot \hat{\bm{z}}),
\end{align}
where $P_2(x) = (3x^2-1)/2$ is a Legendre polynomial and we have assumed that the transition dipole moment is oriented along $\hat{\bm{z}}$, one of the tight tweezer directions. 
It is worth noting that the anisotropic dipole-dipole interaction vanishes for isotropic density distributions.
The Gaussian in-tweezer density can be approximated near the origin as a constant isotropic term,
plus anisotropic terms proportional to $x^2$, $y^2$, $z^2$.
Only the anisotropic terms contribute,
and their $r^2$ scaling results in a finite $U_{dd}$ despite the $r^{-3}$ scaling of the interaction.
Evaluating the integral for our parameters, we find $U_{dd}/h = 5$~kHz.
By choosing a pulse duration of at least 500~$\mu$s, we ensure that the dipolar blockade is effective.

For a molecule in a tweezer, the microwave transitions are both shifted and broadened by the tensor part of the AC Stark shift.
The frequency shift, $\Delta f$, is desirable since it provides a mechanism to address molecules in the tweezer without affecting molecules in the bulk gas.
For this to work, we need $\Delta f$ to be at least as large as $U_{dd}/h$. The broadening, $\delta f$, is undesirable because it reduces the fidelity of the microwave $\pi$ pulses, so we would like $\delta f \ll  U_{dd}/h$. The broadening is due to the range of tweezer intensities seen by the trapped molecule, which is proportional to temperature, so the broadening is smaller than the shift by the ratio of the temperature to the trap depth, which is a factor of 50 for our parameters.
The tensor AC shift depends on the choice of rotational, hyperfine and Zeeman states $(j,f,m_f)$ and can be a significant fraction of the scalar shift for some choices. For CaF, we propose the transition $(1,1^+,\pm1) \rightarrow (2,2^+,\pm 1)$\footnote{Each rotational state $j$ has two hyperfine levels with $f=j$. We use a superscript $+$ to denote the one of higher energy} whose tensor shift at 1064~nm is only 0.15\% of the scalar shift. This results in $\Delta f = 7.8$~kHz and $\delta f = 0.15$~kHz, satisfying our requirements.
Our choice of initial $j=1$ state is not the one produced by deep laser cooling schemes~\cite{cheuk:18,caldwell:19} or optical pumping methods~\cite{williams:18}, but is easily reached using a pair of microwave pulses or a Raman transition~\cite{williams:18,burchesky:21}. 

\begin{figure}
\centering
\includegraphics[width=0.45\textwidth]{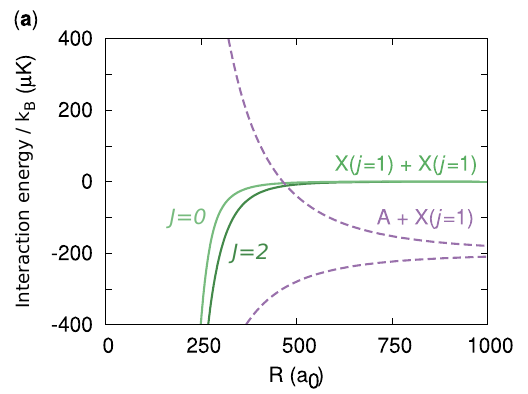}
\includegraphics[width=0.45\textwidth]{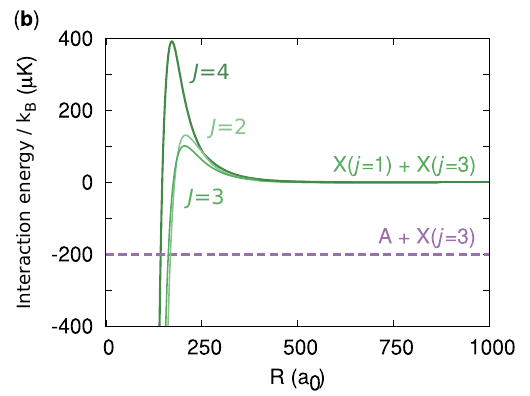}
\caption{
{\bf Molecular interaction potentials}
({\bf a}) Attractive rotational dispersion interaction for two CaF molecules in the $j=1$ rotational state of the electronic ground state (green solid).
Also shown is a schematic of the dressed potentials where one molecule is excited by the cooling light to the $A~^2\Pi$ state, resulting in resonant dipole-dipole interactions (purple dashed). Light-assisted collisions occur at the crossing between the curves.
({\bf b}) Repulsive rotational dispersion interaction for two CaF molecules in the $j=1$ and $j=3$ rotational states \cite{walraven:24a},
which suppresses loss in short-range encounters.
Excitation of the $j=1$ molecule by the cooling light does not produce resonant dipole-dipole interactions in this case, so there are no  light-assisted collisions.
Multiple green lines indicate different possible total angular momentum states as indicated.
\label{fig:collisions}
}
\end{figure}

Let us now turn our attention to collisions.
Figure~\ref{fig:collisions}(a) illustrates the interaction potentials for two $j=1$ molecules in the electronic ground state, shown in green.
If one of the molecules is excited by the cooling light to the $A$ state,
it interacts with the second molecule through resonant dipole-dipole interactions whose strength is proportional to the square of the transition dipole moment.
The field-dressed excited state potential, shown in purple, intersects the ground state and light-assisted collisions occur at the crossing. 
Figure \ref{fig:collisions}(b) illustrates the interaction potentials when the tweezer contains one $j=1$ molecule and one $j=3$ molecule.
The excited state used for laser cooling can only decay to $j=1$ and has no transition dipole moment with $j=3$.
Thus, there are no resonant dipole-dipole interactions and no light-assisted collisions. 

Even in the absence of light-assisted collisions, there can still be collisional losses.
Indeed, ground state ultracold molecules in optical traps typically suffer collisional loss at a rate close to the universal rate \cite{idziaszek:10}, which is $5\times10^{-10}$~cm$^3$/s for CaF. 
Similar loss is observed for two molecules in the $j=1$ rotational state \cite{cheuk:20}.
This universal rate describes loss of molecules at short range,
after passing through an attractive $R^{-6}$ potential.
For two molecules in \emph{different} states that differ in rotational quantum number by more than one, however,
the rotational dispersion interaction is repulsive \cite{walraven:24a}. This produces a repulsive barrier at large distances, as shown in Fig.~\ref{fig:collisions}(b), which can suppress collisional loss by orders of magnitude.

We compute collisional loss rate coefficients as described in more detail in Ref.~\cite{walraven:24a},
for CaF molecules in $|j=1,m\rangle|j'=3,m'\rangle$ initial states.
Our calculations model the molecules as rigid rotors with dipole and quadrupole moments that interact through electrostatic interactions,
as well as through the electronic van der Waals interaction.
We find typical loss rates of $8~(2)\times10^{-12}$~cm$^3$/s,
where the spread in loss rates indicates the $m$, $m'$ dependence.
We also perform exploratory calculations including the electron and nuclear spin,
but find no significant effect on the resulting loss rate coefficients.
Furthermore, we run calculations for SrF and BaF molecules,
where we find comparable loss rates,
in agreement with the weak mass dependence discussed in Ref.~\cite{walraven:24a}.

\begin{figure}
\centering
\includegraphics[width=0.475\textwidth]{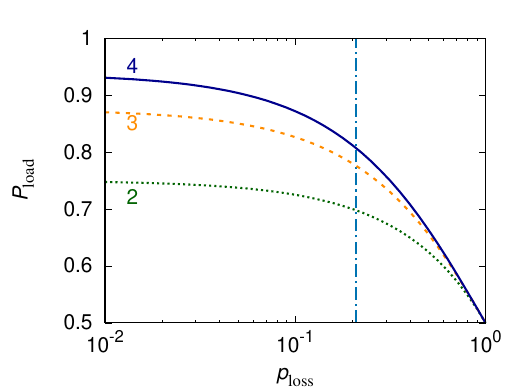}
\caption{
{\bf Performance of the scheme in the presence of residual collisional loss.}
Loading probability as a function of the loss per cycle for 2, 3, and 4~cycles.
The vertical line indicates the probability of loss of CaF molecules during a 10~ms loading cycle when the peak density is $1.5\times10^{13}$~cm$^{-3}$.
\label{fig:loss}
}
\end{figure}

For a $1.5\times10^{13}$~cm$^{-3}$ peak density, the computed loss rate coefficients of a $j=3$ molecule correspond to a collisional lifetime of 25~ms. Choosing a loading time of 10~ms per cycle, and remembering that in the collisional blockade regime the tweezer is empty for half the loading period, the probability of losing a $j=3$ molecule is about 20~\% in each cycle.
The impact of these losses is relatively mild,
since a lost molecule may still be replaced in a subsequent cycle. Figure~\ref{fig:loss} shows the loading efficiency that can be achieved with two, three, or four cycles as a function of the probability of loss per cycle. The vertical line shows the loss probability estimated for the experimental parameters discussed here, which results in approximately $70$~\% loading using two cycles and 80~\% using three or four cycles.

In conclusion, we have proposed a scheme for near-deterministic loading of ultracold molecules into optical tweezers by accumulating at most a single molecule in a rotationally excited storage state for which collisional loss is suppressed.
The required laser cooling into optical tweezers~\cite{anderegg:19},
optical pumping~\cite{williams:18} or coherent population trapping~\cite{cheuk:18,caldwell:19},
and microwave $\pi$ pulses\cite{williams:18} have all been realized experimentally\cite{burchesky:21,holland:23,bao:23,lu:24,bao:24},
though the dipolar blockade and collisional stability in the storage state have yet to be demonstrated.
We estimate our scheme improves the loading efficiency to 80~\% using three cycles.
The success rate is limited by a trade off between a strong dipolar blockade and low residual collisional loss,
which both scale linear with in-tweezer density.
The expected performance is comparable to that for loading of atoms using a scheme that cannot be extended to molecules.
Deterministic loading improves the scalability of quantum computing and simulation based on ultracold molecules in tweezer arrays by limiting the number of rearrangements required in initialization.

\end{document}